\documentclass[12pt,a4paper]{article}
\pdfoutput=1
\usepackage[utf8]{inputenc}
\usepackage{epsfig}
\usepackage{mathtools}
\usepackage{amsfonts}
\usepackage{amssymb}
\usepackage{amsmath}
\usepackage{cancel}
\usepackage{color}
\usepackage{slashed}
\usepackage{cite}
\usepackage{caption}
\usepackage{subcaption}
\usepackage{comment}
\usepackage{marginnote}
\usepackage{accents}

\allowdisplaybreaks

\def\Li              {\mbox{Li}}
\def\la              {\langle}
\def\ra              {\rangle}

\newcommand{\dbtilde}[1]{\accentset{\approx}{#1}}

\def\W{\mathcal{W}}

\def\Q{\mathbb{Q}}
\def\M{\mathbb{M}}
\def\PP{\mathbb{P}}

\def\N4{$\mathcal{N}=4$}

\def\Li              {\mbox{Li}}
\def\la              {\langle}
\def\ra              {\rangle}

\newcommand{\vll}{{\smash{\lambda}}}

\newcommand{\vle}{{\smash{\eta}}}

\begin{document}

\begin{center}

\vspace{1cm}

\renewcommand{\thefootnote}{\fnsymbol{footnote}}

{ \Large\bf Pentagon OPE resummation in \N4 SYM: \\ hexagons with one effective
	particle contribution} \vspace{1cm}

{\large L.V. Bork$^{1,2}$, A.I. Onishchenko$^{3,4}$}~\footnote{E-mail:~\rm{bork@itep.ru},~\rm{onish@theor.jinr.ru}}\vspace{0.5cm}

{\it
$^1$Institute for Theoretical and Experimental Physics, Moscow, Russia,\\
$^2$The Center for Fundamental and Applied Research, \\ All-Russia
Research Institute of Automatics, Moscow, Russia, \\
$^3$Bogoliubov Laboratory of Theoretical Physics, \\ Joint
Institute for Nuclear Research, Dubna, Russia, \\
$^4$Skobeltsyn Institute of Nuclear Physics,  Moscow State University, Moscow, Russia
}

\renewcommand{\thefootnote}{\arabic{footnote}}
\setcounter{footnote}{0}

\vspace{1cm}

\abstract{We present the technique for resummation of flux tube excitations series arising in pentagon operator expansion program for polygonal Wilson loops in \N4 SYM. Here we restrict ourselves with contributions of one-particle effective states and consider as a particular example NMHV$_6$ amplitude at one-loop. The presented technique is also applicable at higher loops  for one effective particle contributions and has the potential for generalization for contributions with more effective particles.
}

\end{center}

\begin{center}
	Keywords: \N4 SYM, amplitudes, pentagon OPE resummation
\end{center}

\newpage

\tableofcontents{}\vspace{0.5cm}

\renewcommand{\theequation}{\thesection.\arabic{equation}}

\section{Introduction}

The discovery of integrability of \N4 SYM in planar limit, see \cite{IntegrabilityReview,IntegrabilityPrimer}  for a review, has led to tremendous progress in our ability to compute different observables in general at arbitrary values of \N4 SYM coupling constant. In particular the collinear OPE or pentagon OPE (POPE) approach to  null-polygonal Wilson loops thanks to duality between amplitudes and (super)Wilson loops \cite{ampWLduality1,ampWLduality2,ampWLduality3,ampWLduality4,ampWLduality5,ampWLduality6}  gives us means for computing scattering amplitudes both at weak and strong values of coupling constant \cite{BubbleAnsatz,YsystemScatteringAmplitudes,OPE_for_W_Loops,Gaiotto:2011dt_OPE,Basso:2013_I,Basso:2013aha_II,Basso:2014koa_III,Basso:2014nra_IV,Basso:2014hfa_All_Helicity_I,Basso:2015rta_All_Helicity_II,Basso:2013vsa_MatrixPart,CordovaEffectiveParticles,BelitskyNMHV-1,BelitskyNMHV-2,BelitskyNMHV-3,AsymptoticBetheAnsatzGKPvacuum,StrongCouplingPOPE-1,StrongCouplingPOPE-2,StrongCouplingPOPE-3,StrongCouplingPOPE-4}. There is also similar approach\footnote{See \cite{IntegrabilityStructureConstants} for introduction.} to structure constants \cite{StructureConstantsPentagons,StructureConstantsWrappingOrder,ClusteringThreePointFunctions} and correlation functions \cite{StructureConstantsPentagons,CorrelationFunctionsIntegrability1,CorrelationFunctionsIntegrability2,CorrelationFunctionsIntegrability3,CorrelationFunctionsIntegrability4,CorrelationFunctionsIntegrability5,CorrelationFunctionsIntegrability6,CorrelationFunctionsIntegrability7,CorrelationFunctionsIntegrability8,CorrelationFunctionsIntegrability9,CorrelationFunctionsIntegrability10,CorrelationFunctionsIntegrability11,CorrelationFunctionsIntegrability12,CorrelationFunctionsIntegrability13,CorrelationFunctionsIntegrability14,CorrelationFunctionsIntegrability15}.

The important problem present within pentagon OPE approach is the problem of resummation of contributions coming from different flux tube excitations. The latter is required if we are going to recover full kinematical dependence (in general kinematics) of scattering amplitudes computed within POPE approach and not restricting ourselves by the collinear limits. At weak coupling  a procedure for resummation of single particle gluon bound states was presented  in \cite{hexagonOPEHarmonicPolylogs,hexagonOPERegge}, see also \cite{Edenresummation} for resummation in the context of $n$-point functions of BPS operators. At strong coupling the procedure for systematic resummation was studied in \cite{AsymptoticBetheAnsatzGKPvacuum,StrongCouplingPOPE-1,StrongCouplingPOPE-2,StrongCouplingPOPE-3,StrongCouplingPOPE-4}, where one should account for resummation of contributions from gluons, scalars, fermions and mesons. On the other hand a systematic approach for resummation at weak coupling \cite{CordovaEffectiveParticles,MHVresummation,BelitskyResummation1,BelitskyResummation2} is tightly connected with the concept of effective particles \cite{Basso:2015rta_All_Helicity_II,CordovaEffectiveParticles}. The latter are formed by fundamental excitations (gluon or its bound states, scalars and large fermions/antifermions) together with arbitrary number of small fermions/antifermions.  The introduction of effective particles allowed to reconstruct several scattering amplitudes in general kinematics at tree level \cite{CordovaEffectiveParticles,BelitskyResummation1,BelitskyResummation2} and MHV hexagon amplitude at one-loop level. Here we are going to extend these results and present the technique for resummation of one effective particle contributions to hexagon amplitudes at arbitrary order of perturbation theory. As a particular example we consider
NMHV$_6$ amplitude at one-loop. These technique has the potential for generalization for both higher point scattering amplitudes and contributions with more then one effective particle.

This paper is organized as follows. In section 2 we give a brief introduction to the collinear pentagon OPE approach  and the concept of effective particles.  Section 3 contains details of our resummation technique in the case of NMHV one-loop hexagon together with the prescription for treating one effective particle contributions in the case of hexagons at arbitrary loop order.  Finally in section 4 we come with our conclusion. The appendices contain explanation of notation together with different details of our calculation of NMHV hexagon amplitude at one-loop.

\section{Hexagon pentagon OPE: one effective particle states}

Let us briefly remind the essential for our further discussion ideas and facts about pentagon OPE (POPE) approach and effective particle concept. For detailed discussion see \cite{Basso:2013_I,Basso:2013aha_II,Basso:2014koa_III,Basso:2014nra_IV} and \cite{Basso:2014hfa_All_Helicity_I,Basso:2015rta_All_Helicity_II,Basso:2013vsa_MatrixPart,CordovaEffectiveParticles}. Using duality between amplitudes and (super)Wilson loops \cite{ampWLduality1,ampWLduality2,ampWLduality3,ampWLduality4,ampWLduality5,ampWLduality6} one can recast the problem of calculation of components of the finite remainder function $\mathcal{R}_n^{(k)}$ for the $\mbox{N}^k\mbox{MHV}_n$ amplitude\footnote{See appendix A for more details.} into the problem of evaluation of the ratios $\mathcal{W}_n$ of vacuum expectation values of polygonal lightlike Wilson loops with fields inserted on edges and cusps \cite{superWL1,superWL2}. The latter within pentagon OPE approach are then decomposed into $n-3$ successive fluxes/squares  \cite{OPE_for_W_Loops,Gaiotto:2011dt_OPE,Basso:2013_I}. The first essential ingredient for building pentagon OPE expansion is given by the knowledge of color flux tube excitation spectrum, which in \N4 SYM is known thanks to integrability for \emph{arbitrary} values of coupling constant $g$ \cite{Basso:2010in_GKP_string}. The second important ingredient is supplied by transitions from one flux to another induced by pentagon operators. The matrix elements of the latter could be also computed at any coupling using integrable bootstrap \cite{Basso:2013_I}, see further development in  \cite{Basso:2013aha_II,Basso:2014koa_III,Basso:2014nra_IV,Basso:2014hfa_All_Helicity_I,Basso:2015rta_All_Helicity_II}

To be more specific, the renormalized\footnote{See \cite{Basso:2013_I} for more details.} vacuum expectation value of  $n$ polygonal super Wilson loop within pentagon OPE approach is given by \cite{Basso:2014hfa_All_Helicity_I,Basso:2015rta_All_Helicity_II}:
\begin{equation}
\W_n = \sum_{\Psi_i}\PP (0|\Psi_1)\PP (\Psi_1|\Psi_2)\ldots \PP (\Psi_{n-6}|\Psi_{n-5})\PP (\Psi_{n-5}|0)e^{\sum_j (-E_j\tau_j + i p_j\sigma_j + i m_j\phi_j)} ,
\end{equation}
where $\{\tau_i, \sigma_i, \phi_i\}$ is a base of conformal ratios, parameterizing propagation of $\Psi_i$ excitation (in general multi-particle) in $i$-th flux/square. The $E_i$, $p_i$ and $m_i$ denote energy\footnote{The energies of excitations are in one to one correspondence with anomalous dimensions of corresponding single trace GKP operators \cite{GKPstring} $Tr(ZD^{S_1}\mathcal{O}D_{+}^{S_2}Z)$, where $Z$ is one of three complex scalars in \N4 SYM, $D_{+} = n_{+}^{\mu}D_{\mu} = D_0+D_3$ is the light-cone covariant derivative and $\mathcal{O}$ is some monomial constructed from $\{F_b,\psi,\phi\}$ fields. It is also assumed, that $S_1+S_2 \gg 1$. }, momentum and angular momentum (helicity) of $i$-th excitation. The  transition probabilities from one flux to another $\PP(\Psi_i|\Psi_j)$ are described by matrix elements of  charged or super pentagon operators introduced in \cite{Basso:2014hfa_All_Helicity_I}:
\begin{equation}\label{superPentagon}
\PP = P + \chi^A P_A + \chi^A\chi^B P_{AB} + \chi^A\chi^B\chi^C P_{ABC} + \chi^A\chi^B\chi^C\chi^D P_{ABCD}\, ,
\end{equation}
where $\chi^A$ is a Grassmann parameter transforming in the fundamental representation of $SU(4)_R$ $R$-symmetry group. $P_{A_1\ldots A_k}$ or $P^{[k]}$ for short is charged pentagon transition transforming as $k$-th antisymmetric product. Charged\footnote{When thought in terms of pentagon polygonal Wilson loops the charged pentagons have additional fields insertions at their cusps and edges compared to usual uncharged pentagons.} pentagon transitions contrary to ordinary uncharged pentagon transitions $P$ used to describe MHV amplitudes via bosonic polygonal Wilson loops may produce states with non-zero $R$-charge. For example, the creation amplitude $P_{AB}(0|\ldots)$ may produce scalar fields $\phi_{AB}$ out of the vacuum, as the quantum numbers of the latter match those of pentagon.

In a particular case of hexagon the pentagon OPE expansion gives\footnote{$1/S_m$ is a symmetry factor.} \cite{Basso:2015rta_All_Helicity_II,CordovaEffectiveParticles}:
\begin{equation}
\W_6^{[r_1,r_2]} = \sum_m \frac{1}{S_m}\int\frac{du_1\ldots du_m}{(2\pi)^m} \Pi_{dyn}\times\Pi_{FF}^{[r_1,r_2]}\times\Pi_{mat}^{[r_1,r_2]}, \label{W6r1r2}
\end{equation}
where $u_k$ are rapidities of intermediate particle states and  $r_1$, $r_2$ are $SU(4)_R$ charges of top and bottom pentagons.  The latter, as was mentioned above, are related to the particle content of the $\mathcal{R}_6$ remainder function. In the
NMHV case $r_1,r_2$ are constrained, such that $r_1+r_2=4$ and as a consequence NMHV hexagon has five different POPE components. The multi-particle flux tube excitations are build from {\it fundamental excitations} represented\footnote{Here, we suppressed $SU(4)_R$ and projected Lorenz indexes of fields.} by   gluon bound states, fermions, antifermions and scalars \cite{Basso:2010in_GKP_string}: $\{F_b, \psi, \bar{\psi}, \phi\}$. The integrand in Eq.\eqref{W6r1r2} has a factorized form and consists from coupling dependent {\it dynamical} $\Pi_{dyn}$ and {\it form factor}  $\Pi_{FF}^{[r_1,r_2]}$ parts. The {\it matrix} part $\Pi_{mat}^{[r_1,r_2]}$, which takes into account $SU(4)_R$ structure of flux excitations, on the other hand is coupling independent.

The dynamical part contribution has the form
\begin{equation}
\Pi_{dyn} = \prod_j \mu (u_j) e^{-E(u_j)\tau + ip(u_j)\sigma + im_j\phi}\times \prod_{i<j}\frac{1}{|P(u_i|u_j)|^2} , \label{Pidyn}
\end{equation}
where $\tau$, $\sigma$ and $\phi$  are real parameters encoding \emph{all} external kinematical dependence (they parameterize three conformal cross ratios $u_1,u_2,u_3$ on which $\W_6^{[r_1,r_2]}$ depends) of the problem. They also have the meanings of flux tube time, space and angle coordinates respectively. In addition, the $\tau$ variable parametrizes the measure of collinearity of two adjacent amplitude momenta with the limit $\tau \rightarrow \infty$ corresponding to collinear configuration \cite{OPE_for_W_Loops,Basso:2013_I}, see appendix \ref{Dixonresult} for more details. $P(u_i|u_j)$ in Eq.\eqref{Pidyn} are uncharged pentagon transitions\footnote{These functions depend only on the types of fundamental excitations, their spectral parameters and coupling constant $g$.} between different fundamental excitations and $\mu (u_i)$ are corresponding measures. The expressions for $P(u_i|u_j)$ and $\mu(u_i)$ are known for {\it arbitrary} values of coupling constant and can be found in \cite{Basso:2013_I,Basso:2013aha_II,Basso:2014koa_III,Basso:2014nra_IV}.

The form factor part contribution is obtained by expressing charged pentagon transitions in terms of uncharged ones and is nontrivial only for NMHV hexagons. In our case it is given by \cite{Basso:2015rta_All_Helicity_II,CordovaEffectiveParticles}:
\begin{eqnarray}
\Pi_{FF}^{[r_1,r_2]}=g^{\frac{1}{8}r_1(r_1-4)+\frac{1}{8}r_2(r_2-4)}\times\prod_ih(u_i)^{r_1-r_2},
\end{eqnarray}
where $h(u_i)$ are the so called form factors and are also known for arbitrary values of coupling constant \cite{Basso:2015rta_All_Helicity_II}.

The matrix part contribution takes into account contraction of $SU(4)_R$ indexes of each pentagon and in our case takes the form of the integral over auxiliary roots \cite{HexagonsFiniteCoupling}:
\begin{eqnarray}
\Pi_{mat}^{[r_1,r_2]}&=&\frac{1}{K_1!K_2!K_3!}\int
\prod\limits_{i=1}^{K_1}\frac{dw_i^1}{2\pi}
\prod\limits_{i=1}^{K_2}\frac{dw_i^2}{2\pi}
\prod\limits_{i=1}^{K_3}\frac{dw_i^3}{2\pi}\times\\
&\times&\frac{g({\bf w}^1)g({\bf w}^2)g({\bf w}^3)}{f({\bf w}^1,{\bf w}^2)f({\bf w}^2,{\bf w}^3)f({\bf w}^1,{\bf v})f({\bf w}^2,{\bf s})f({\bf w}^3,{\bf \bar v})},
\end{eqnarray}
where $w_i$ are auxiliary roots (rapidities) corresponding to three nodes of $SU(4)$ Dynkin diagram and $\{v_i, s_i, {\bar v}_i\}$ are rapidities for fermions, scalars and antifermions correspondingly. In addition,
$g({\bf w})=\prod_{i<j}(w_i-w_j)^2[(w_i-w_j)^2+1]$ and $f({\bf w},{\bf v})=\prod_{i,j}[(w_i-v_j)^2+\frac{1}{4}]$. The number of auxiliary rapidities $K_1$, $K_2$ and $K_3$ are solutions of the following system of equations:
\begin{eqnarray}
N_\psi -2 K_1 +K_2  &=& \delta_{r_1,3}\,\\
N_\phi + K_1- 2 K_2 + K_3 &=& \delta_{r_1,2}\,\\
N_{\bar\psi}\,\,\, + K_2 -2K_3 &=& \delta_{r_1,1}\,,
\end{eqnarray}
where $N_\psi$, $N_\phi$ and $N_{\bar\psi}$ are respectively the number of fermions, scalars and antifermions in multi-particle excitation.

In the weak coupling regime within POPE approach to hexagon amplitudes contributions of different excitations scale as $g^{2l}e^{-\tau N}$, where $N$ is the total twist of corresponding multi-particle state and $l$ - number of loops. For such expansion to be convergent one has to consider only collinear enough configurations of momenta with $\tau > 1$. The coefficients in front of $g^{2l}e^{-\tau N}$ could be compared with independently computed amplitude results expanded in collinear limit \cite{Basso:2013aha_II,Basso:2014koa_III,hexagonOPEHarmonicPolylogs}. They can also serve as predictions for such collinear limits \cite{hexagonOPEHarmonicPolylogs,hexagonOPERegge}. On the other hand,
one can try to re-sum contributions of all possible excitations contributing at a given loop order $l$. Together with analytical continuation of resummation result  to $\tau \leq 1$ this should allow for a full reconstruction of the whole kinematical dependence \cite{Basso:2014nra_IV}. The possibility of such resummation also implies means of getting POPE results for  $\mathcal{R}_n$ remainder functions without any reference to \N4 lagrangian and corresponding Feynman rules or unitarity cuts.

To make such resummation possible one has to understand the hierarchy of flux tube excitations in the weak coupling regime. That is, we need to know when and which excitation starts to give contribution to perturbative expansion. The useful hint comes from the structure of fermionic excitations, which are separated into {\it large} $\psi$ and {\it small} $\psi_s$ fermions. The latter property is due to the fact, that in terms of Bethe rapidity the fermionic excitations are defined on two-sheeted Riemann surface \cite{Basso:2014koa_III}. On one Riemann sheet the fermion momentum is large, while on the other it is small. When attached to another particle, small fermions $\psi_s$, $\bar{\psi}_s$ act as a supersymmetry generators \cite{Alday:2007mf}. The action  of $\psi_s\bar{\psi}_s$ pairs (or derivatives $D_{+}$) creates $SL(2)$ conformal {\it descendants} as at weak coupling there is an enhancement of symmetry\footnote{This symmetry is exact only at one loop level, however the same bookkeeping turns out to be useful also at higher loops.} from $SU(4)$ to $SL(2|4)$ \cite{Gaiotto:2010fk,Gaiotto:2011dt_OPE,Basso:2010in_GKP_string}

The very useful notion for the purposes of pentagon OPE resummation is provided by the concept of effective particles \cite{Basso:2015rta_All_Helicity_II,CordovaEffectiveParticles}. By effective particle we will understand a fundamental excitation together with  arbitrary number ("sea") of small fermion (antifermion) excitations $N_{\psi_s}$ ($N_{\bar{\psi}_s}$). Having more then one fundamental excitation surrounded by the sea of small fermions/antifermions will lead to more then one effective particle state. Integrating out small fermion/antifermion rapidities together with auxiliary $SU(4)_R$ roots leads to the description of effective particles in terms of Bethe string complexes. In general, the effective particle (excitation) is described by three parameters: the helicity or angular momentum of excitation $a$, its descendant number $n$  and $SU(4)_R$ representation in which it transforms. One can show that for the $\mbox{NMHV}_6$ amplitude the contribution of one effective particle is sufficient for its reconstruction both at tree and one-loop (LO and NLO) levels. The account for two effective particles is enough to reconstruct two, three and four loops\footnote{For $\mbox{MHV}_6$ amplitude one effective particle is sufficient for one-loop reconstruction \cite{MHVresummation} and combination of one and two effective particles will be enough to reconstruct amplitude up to (including) five loops \cite{Basso:2015rta_All_Helicity_II,CordovaEffectiveParticles}.}. So, we see that the number of effective particles we should take into account grows rather slowly with loop order.

To demonstrate our resummation technique in the next section we will use $\W_6^{[2,2]}$ NMHV POPE component. In this case, restricting ourselves with one effective particle contributions,  we should account for the following effective particles, transforming in vector representation of $SU(4)_R$ \cite{CordovaEffectiveParticles}:
\begin{gather}
\Phi^{1}_{a,n} = F_a\psi_s\psi_s(\bar{\psi}_s\psi_s)^n , \quad
\Phi^{2}_{a,n} = \bar{F}^a\bar{\psi}_s\bar{\psi}_s(\bar{\psi}_s\psi_s)^n , \nonumber \\
\Phi^{3}_{n} = \phi(\bar{\psi}_s\psi_s)^n , \quad \Phi^{4}_{n} = \psi\psi_s(\bar{\psi}_s\psi_s)^n , \quad \Phi^{5}_{n} = \bar{\psi}\bar{\psi}_s(\bar{\psi}_s\psi_s)^n .
\end{gather}
In the case $n=0$ the above effective particles are $SL(2)$ conformal primaries.  Taking integrals over small fermion/antifermion rapidities and auxiliary $SU(4)_R$ roots by residues the expression for $\W_6^{[2,2]}$ POPE component takes the form \cite{CordovaEffectiveParticles}:
\begin{equation}\label{OPE_eff_particle}
\W_6^{[2,2]} = \sum_{\Phi}\int\frac{du}{2\pi} e^{E_{\Phi}(u)\tau + i p_{\Phi}(u)\sigma + i m_{\Phi}\phi}\mu_{\Phi}^{[2,2]}(u) +\ldots,
\end{equation}
where $\ldots$ stands for multiple effective particle contributions and the expressions for energies $E_{\Phi}(u)$, momenta $p_{\Phi}(u)$, angular momenta  $m_{\Phi}$  and integration measures $\mu_{\Phi}^{[2,2]}(u)$ of effective particles can be found in Appendix \ref{MeasureEnergiesMomenta}.

The first steps to resummation of series in Eq. \eqref{OPE_eff_particle} were made in \cite{CordovaEffectiveParticles} at LO. However, we found that the method employed there is somewhat hard to generalize to higher orders of perturbation theory. So in the following section we are going to present an algorithm which
should allow one to compute series representation for $\W_6^{[r_1,r_2]}$ functions similar to that for $\W_6^{[2,2]}$ \eqref{OPE_eff_particle}  in terms of multiple polylogarithms \cite{multiple-polylogs1,multiple-polylogs2} of kinematical variables at any order of perturbation theory. Presumably, the same algorithm should be also applicable to other cases with $n >6$ and contributions with more effective particles.
As an illustration for our method we will consider LO and NLO contributions to $\W_6^{[2,2]}$ POPE component. In this case it is sufficient to consider one effective particle contributions only. To compare the results of pentagon OPE resummation with results for hexagon amplitudes computed with other methods we should recall that the usual way to package together all helicity amplitudes is to use super Wilson loop \cite{superWL1,superWL2}:
\begin{equation}
\mathbb{W}_6 = W_{6, MHV} + \eta_i^1\eta_j^2\eta_k^3\eta_l^4 W_{6, NMHV}^{\la ijkl\ra} + \ldots , \label{superWL6}
\end{equation}
where $W_{NMHV}$ is the NMHV amplitude divided by Parke-Taylor MHV factor. Here, the Grassmann variables $\eta_j^A$ are Grassmann components of hexagon momentum twistors (see Appendix \ref{Dixonresult} for more details) with upper index transforming in the fundamental representation of $SU(4)_R$ and lower index labeling the edge of hexagon. The important thing here is that these Grassmann variables are different from those used within POPE framework \eqref{superPentagon}. Nevertheless there is a map from one set of Grassmann variables to another \cite{Basso:2014hfa_All_Helicity_I}. In particular, it turns out that \cite{Basso:2014hfa_All_Helicity_I,CordovaEffectiveParticles}:
\begin{equation}
\W_6^{[2,2]} = -\W_6^{\la 1144\ra} ,
\end{equation}
where $\W_6^{\la 1144\ra}$ is the $W_{6, NMHV}^{\la ijkl\ra}$  component from Eq.\eqref{superWL6}.

\section{Resummation technique}

Before presenting the general algorithm for treating one effective particle contributions in the case of hexagons, let us first start with the particular example of hexagon NMHV amplitude and  later formulate the general prescription for the resummation of one effective particle contributions to hexagon Wilson loops in \N4 SYM. Up to one loop the expression for $\W_6^{\la 1 1 4 4\ra}$ component takes the  form\footnote{See appendix \ref{MeasureEnergiesMomenta} for the expression for $\W_6^{[2,2]}$, which we expand up to one-loop order. We have also made change of variables $u\to -i u$, so that now the integration contour goes along imaginary axis. Also $g\equiv g_{YM}^2N_c/(16\pi^2)$.}:
\begin{align}
\mathcal{W}_6^{\la 1 1 4 4\ra} &= -\W_6^{[2,2]} =
 \sum_{a=-\infty}^{\infty}\sum_{n=0}^{\infty}
\int\frac{d u}{2\pi i} e^{-(|a|+2n+1)\tau + 2 u\sigma + i a\phi} (-1)^{a+n}  \nonumber \\ & \times
\frac{\Gamma \left(\frac{|a|}{2}-u-\frac{1}{2}\right)\Gamma\left(
\frac{|a|}{2}+u+\frac{3}{2}+n	
\right)^2}{\Gamma\left(\frac{|a|}{2}+u+\frac{3}{2}\right)\Gamma \left(
|a| + n + 1
\right) n!}\left\{
1 + g^2 f_{a,n}^{NLO}(u) + \mathcal{O}(g^4) \label{W6_1144_start}
\right\}\, ,
\end{align}
where
\begin{multline}
f_{a,n}^{NLO}(u) = \frac{\pi^2}{3} - \frac{6}{(1-|a|+2u)^2} + \frac{2}{|a|(1-|a|+2u)} + \frac{2}{(1+|a|+2u)^2} - \frac{2}{|a|(1+|a|+2u)}  \\
-2\tau \left[
2\gamma_E + \Psi^{(0)}\left(\frac{|a|+1}{2}-u\right)
+ \Psi^{(0)}\left(\frac{|a|+1}{2}+u\right)
\right]  \\
 +2\sigma \left[
\Psi^{(0)}\left(
\frac{|a|-1}{2}-u
\right) + \Psi^{(0)}\left(
\frac{|a|+3}{2} + u
\right) - 2 \Psi^{(0)}\left(
\frac{|a|+3}{2} + n + u
\right)
\right] \\
- \frac{1}{2}\left[
2\gamma_E + \Psi^{(0)}\left(
\frac{|a|+1}{2}-u
\right) + \Psi\left(
\frac{|a|+1}{2}+u
\right)
\right]^2  \\
 -\frac{1}{2}\left[
\Psi^{(0)}\left(
\frac{|a|-1}{2} - u
\right) + \Psi^{(0)}\left(
\frac{|a|+3}{2} + u
\right) - 2\Psi^{(0)}\left(
\frac{|a|+3}{2} + n + u
\right)
\right]^2  \\
 -\Psi^{(1)}\left(
\frac{|a|+1}{2} - u
\right) - \Psi^{(1)}\left(
\frac{|a|+1}{2} + u
\right) + 2 \Psi^{(1)}\left(
\frac{|a|+3}{2}+u
\right) - 2 \Psi^{(1)}\left(
\frac{|a|+3}{2} + n + u
\right)\, ,
\end{multline}
Here $\Psi^{(n)}(z)$ are polygamma functions. To evaluate the above expression both at LO and higher we start with taking residues in $u$-variable. To achieve this we first use reflection  identities
\begin{align}
\Gamma \left(\frac{|a|-1}{2}-u\right) &= \frac{\pi\csc\left(
\frac{\pi (|a|-1)}{2} - \pi u 	
\right)}{\Gamma\left(
\frac{3-|a|}{2}+u
\right)}\, , \label{reflection-identity-1}\\
\Psi^{(n)}\left(
\frac{|a|+3}{2}-u
\right) &= (-1)^n \Psi^{(n)}\left(
u - \frac{|a|+1}{2}
\right) - \pi\frac{\partial^n}{\partial u^n}\cot\left(
\frac{\pi (|a|+3)}{2} - \pi u
\right)  \label{reflection-identity-2}
\end{align}
to isolate singular terms into elementary functions with known Taylor expansions. It is also convenient to transform present polygamma functions to the same argument as far as possible using the following recurrence relation
\begin{align}
\Psi^{(n)}(z+1) = \Psi (z) + (-1)^n n! z^{-n-1}\, . \label{Psitransform}
\end{align}
Note, that it is the general procedure when taking Mellin-Barnes integrals and was used already in the  context of collinear OPE in \cite{hexagonOPEHarmonicPolylogs,hexagonOPERegge}. Now, taking residues at $u=\frac{|a|-1}{2}+k$ we get:
\begin{equation}
\mathcal{W}_6^{\la 1 1 4 4\ra} = \mathcal{W}_{6,m}^{\la 1 1 4 4\ra} + \mathcal{W}_{6,b}^{\la 1 1 4 4\ra}\, ,
\end{equation}
where subscripts $m$ and $b$ denote what we call main and boundary\footnote{The boundary contributions start contributing from NLO order.} contributions. The latter are given by:
\begin{multline}
\mathcal{W}_{6,m}^{\la 1 1 4 4\ra} =
\sum_{a=-\infty, a\neq 0}^{\infty}\sum_{n=0}^{\infty}\sum_{k=0}^{\infty}
\frac{(-1)^{a+n+k}}{k!n!}e^{-(|a|+2n+1)\tau + (|a|+2k-1)\sigma + i a \phi} \\
\times \frac{(|a|+n+k)!}{(|a|+k)!}\frac{(|a|+n+k)!}{(|a|+n)!}\left\{
1 +  g^2 \tilde{f}_{a,n}^{NLO}(k) + \mathcal{O}(g^4)
\right\} \\
+ \sum_{n=0}^{\infty}\sum_{k=1}^{\infty} \frac{(-1)^{n+k}}{k!n!}
e^{-(2n+1)\tau + (2k-1)\sigma}\frac{(n+k)!}{k!}\frac{(n+k)!}{n!}\left\{
1 +  g^2 \tilde{f}_{0,n}^{NLO}(k) + \mathcal{O}(g^4)
\right\}\, ,
\end{multline}
and
\begin{multline}
\mathcal{W}_{6,b}^{\la 1 1 4 4\ra} = \sum_{a=-\infty, a\neq 0}^{\infty}\sum_{n=0}^{\infty} \frac{(-1)^{a+n} }{n!}e^{-(|a|+2n+1)\tau + (|a|-1)\sigma + i a \phi}\\
\times \frac{(|a|+n)!}{|a|!}\left\{
g^2 \dbtilde{f}_{a,n}^{NLO} + \mathcal{O}(g^4)
\right\}
\end{multline}
where
\begin{multline}
\tilde{f}_{a,n}^{NLO} (k) = \frac{\pi^2}{3} - \frac{2}{k^2}(1-\delta_{k,0}) - \frac{2}{(k+|a|)^2} + \frac{2(\sigma + \tau)}{k}(1-\delta_{k,0}) + \frac{2(\sigma + \tau)}{k+|a|} - 4\sigma\tau \\
-2\Psi^{(1)}(|a|+n+k+1) -2 \Big(
\Psi^{(0)}(|a|+n+k+1) + \gamma_E
\Big)^2
\\ + 2\left(
\frac{1}{k}(1-\delta_{k,0}) + \frac{1}{k+|a|} - 2\sigma - 2\tau
\right)\Big(
\Psi^{(0)}(|a|+n+k+1) + \gamma_E
\Big)\, . \label{fmainNLO}
\end{multline}
and
\begin{multline}
\dbtilde{f}_{a,n}^{NLO} = 4\sigma\tau - \frac{\pi^2}{6} + \Psi^{(1)}(1+|a|) + 4\tau \Big(
\Psi^{(0)}(|a|+n+1) + \gamma_E
\Big) \\  - \Big(
\Psi^{(0)}(1+|a|)+\gamma_E
\Big)^2  + 2\Big(
\Psi^{(0)}(|a|+1)+\gamma_E
\Big)\Big(
\sigma - \tau + \Psi^{(0)}(|a|+n+1)
\Big)\, . \label{fboundaryNLO}
\end{multline}
Introducing notations $x=e^{-\tau}, y=e^{\sigma}, z=e^{i\phi}$ the above expressions take the form
\begin{multline}
\mathcal{W}_{6,m}^{\la 1 1 4 4\ra} = \sum_{a=1}^{\infty}\sum_{n=0}^{\infty}\sum_{k=0}^{\infty} (-1)^{a+n+k}
x^{a+2n+1}y^{a+2k-1}\left(z^a+z^{-a}\right) \\ \times
\binom{a+n+k}{n} \binom{a+n+k}{k} \left\{
1 +  g^2 \tilde{f}_{a,n}^{NLO}(k) + \mathcal{O}(g^4)
\right\} \\
+ \sum_{n=0}^{\infty}\sum_{k=1}^{\infty} (-1)^{n+k} x^{2n+1} y^{2k-1}\binom{n+k}{n}\binom{n+k}{k} \left\{
1 +  g^2 \tilde{f}_{0,n}^{NLO}(k) + \mathcal{O}(g^4)
\right\}
\end{multline}
and
\begin{multline}
\mathcal{W}_{6,b}^{\la 1 1 4 4\ra} = \sum_{a=1}^{\infty}\sum_{n=0}^{\infty} (-1)^{a+n}
x^{a+2n+1}y^{a-1}\left(z^a+z^{-a}\right)
\binom{a+n}{n}  \left\{
 g^2 \dbtilde{f}_{a,n}^{NLO} + \mathcal{O}(g^4)
\right\}
\end{multline}

\subsection{LO}

To evaluate the sums left after taking residues in $u$ it is convenient to introduce the following integral representations for binomial coefficients\footnote{Here, the integration contour is actually going around $z=0$.}:
\begin{equation}
\binom{n}{k}  = \frac{1}{2\pi i}\int_{|t|=1}(t+1)^n t^{-k-1} dt\, . \label{binomial-integral-rep}
\end{equation}
Then at leading order we have
\begin{multline}
\mathcal{W}_6^{\la 1 1 4 4\ra, LO} = \sum_{a=1}^{\infty}\sum_{n=0}^{\infty}\sum_{k=0}^{\infty}\frac{(-1)^{a+n+k}}{(2\pi i)^2}\int_{|t_1|=1}dt_1\int_{|t_2|=1}dt_2 \\
\times  x^{a+2n+1} y^{a+2k-1} (z^a+z^{-a})\left[
(t_1+1)(t_2+1)
\right]^{a+n+k}t_1^{-n-1}t_2^{-k-1} \\
+\sum_{n=0}^{\infty}\sum_{k=1}^{\infty}\frac{(-1)^{n+k}}{(2\pi i)^2}\int_{|t_1|=1}dt_1\int_{|t_2|=1}dt_2  x^{2n+1} y^{2k-1}
\left[
(t_1+1)(t_2+1)
\right]^{a+n+k}t_1^{-n-1}t_2^{-k-1}
\end{multline}
Now, the series summation is straightforward and we get
\begin{multline}
\mathcal{W}_6^{\la 1 1 4 4\ra, LO} = -\frac{1}{(2\pi i)^2}\int_{|t_1|=1}dt_1\int_{|t_2|=1}dt_2 \frac{(1+t_1)(1+t_2)x}{(t_1+(1+t_1)(1+t_2)x^2)(t_2+(1+t_1)(1+t_2)y^2)} \\
\times\left\{
\frac{y}{t_2} + \frac{x}{(1+t_1)(1+t_2)xy+z} + \frac{xz}{1+(1+t_1)(1+t_2)xyz}
\right\}\, .
\end{multline}
Next, performing partial fractioning in $t_2$ variable and taking residues at $t_2=0$ and $t_2=-1+\frac{1}{1+(1+t_1)y^2}$ together with subsequent residues in $t_1$ at $t_1 = -\frac{x^2}{1+x^2}$ and $t_1 = \frac{-1-x^2-y^2+\sqrt{(1+x^2+y^2)^2-4x^2y^2}}{2y^2}$ we get
\begin{equation}
\mathcal{W}_6^{\la 1 1 4 4\ra, LO} = \frac{x}{y}\left(
\frac{z}{z+(y+xz)(x+yz)} - \frac{1}{1+x^2}\,
\right)
\end{equation}
in agreement with \cite{CordovaEffectiveParticles}. We would like to clarify the particular choice of points, at which residues over $t_1$ and $t_2$ should be taken. First, we know that in the limit $x\to 0, y\to 0$ the residue should be taken at the point $t_1 = t_2 = 0$ and so our points at which we took residues should go to this particular point in this limit. And of course we may greatly benefit from numeric checks for some particular values of Mandelstam variables that we actually get the correct expression in the end.

\subsection{NLO}

The integration procedure at NLO and higher goes similar to the LO case. To illustrate the presented technique let us consider evaluation of several terms in NLO contribution. The results for the rest of terms could be found in accompanying {\it Mathematica} notebook. The different terms in the main contribution at NLO can be written as
\begin{multline}
\mathcal{W}_{6,m}^{\la 1 1 4 4\ra}\Big[ f_{a,n}(k) \Big] = \sum_{a=1}^{\infty}\sum_{n=0}^{\infty}\sum_{k=0}^{\infty} (-1)^{a+n+k}
x^{a+2n+1}y^{a+2k-1}\left(z^a+z^{-a}\right) \\ \times
\binom{a+n+k}{n} \binom{a+n+k}{k} f_{a,n}(k)  \\
+ \sum_{n=0}^{\infty}\sum_{k=1}^{\infty} (-1)^{n+k} x^{2n+1} y^{2k-1}\binom{n+k}{n}\binom{n+k}{k} f_{0,n}(k)
\end{multline}
where $f_{a,n}(k)$ are given by terms in the sum of Eq. \eqref{fmainNLO}. To calculate the latter it is convenient to express $1/k^n$ and $1/(k+a)^n$ factors in terms of polygamma functions as
\begin{equation}
\frac{1}{z^n} = \frac{(-1)^n}{(n-1)!}\left[
\Psi^{(n-1)}(z) - \Psi^{(n-1)}(z+1)
\right] \label{simple-fractions-polygamma}
\end{equation}
and use for polygamma functions the following integral representations
\begin{align}
\Psi^{(n)}(z) &= \int_0^1 \frac{x^{z-1}\log^n x}{x-1}dx\,  \quad\mbox{ if } n > 0\, , \\
\Psi^{(0)}(z) &= \int_0^1 \frac{1-x^{z-1}}{1-x}dx - \gamma_E\, .
\end{align}
For example following the above prescription for $1/(k+a)$ term, that is rewriting
\begin{equation}
\frac{1}{k+a} = \Psi^{(0)} (k+a+1) - \Psi^{(0)} (k+a) = \int_0^1 dx_1 x_1^{k+a-1}
\end{equation}
and using the same integral representations for binomial coefficients as at LO we may again easily sum the geometric series in $a$, $n$ and $k$ and get
\begin{multline}
\mathcal{W}_{6,m}^{\la 1 1 4 4\ra}\Big[\frac{1}{k+a}\Big] = \int_0^1 dx_1\int_{|t_1|=1}dt_1 \int_{|t_2|=1}dt_2 \\
\frac{x (1+t_1)(1+t_2)}{(t_1+(1+t_1)(1+t_2)x^2)(t_2+(1+t_1)(1+t_2)x_1 y^2)}\\
\times \left\{
\frac{y}{t_2} + \frac{x}{(1+t_1)(1+t_2)x y x_1 + z} + \frac{x z}{1+(1+t_1)(1+t_2)x y z x_1}
\right\} .
\end{multline}
Now, taking residues in $t_2$ at $t_2 = 0$ and $t_2 = -1 + \frac{1}{1+(1+t_1)y^2x_1}$ together with subsequent residues in $t_1$ at $t_1 = -\frac{x^2}{1+x^2}$ and $t_1 = \frac{-1-x^2-y^2 x_1 + \sqrt{(1+x^2+y^2x_1)^2-4x^2y^2x_1}}{2y^2x_1}$ we have
\begin{multline}
\mathcal{W}_{6,m}^{\la 1 1 4 4\ra}\Big[\frac{1}{k+a}\Big] = \int_0^1 dx_1
\frac{x}{2yx_1}\Bigg\{ \frac{2}{1+x^2} \\
+ \frac{xyz(x_1-1)(2xyz(1+x_1)+ (1+x^2+y^2x_1)(1+z^2))}{p_1(x,y,x_1)(z+(xz+yx_1)(x+yz))(z+(y+xz)(x+yzx_1))} \\
-\frac{z (2zx^2+2z(1+y^2x_1)+xy(1+z^2)(1+x_1))}{(z+(xz+yx_1)(x+yz))(z+(y+xz)(x+yzx_1))} \Bigg\}\, ,
\end{multline}
where $p_1(x,y,x_1) = \sqrt{(1+x^2+y^2x_1)^2-4x^2y^2x_1}$. Note, that the points at which residues were taken are deformations of corresponding points we had at LO for $x_1 = 1$. Next, the integral in $x_1$ could be easily evaluated by rationalizing root in $p_1(x,y,x_1)$ with the following variable substitution
\begin{equation}
x_1 = \frac{2x^2}{y^2-t} - \frac{2}{y^2+t}
\end{equation}
As a result we get
\begin{multline}
\mathcal{W}_{6,m}^{\la 1 1 4 4\ra}\Big[\frac{1}{k+a}\Big] = \frac{x}{y (1+x^2)}\Bigg\{ 2\log\left(\frac{2xy^2}{1+x^2}\right) - \log \left(
1+x^2+y^2-p(x,y)
\right) \\ - \log\left(-1-x^2+y^2+p(x,y)\right)\Bigg\} + \frac{x}{y (1+x^2+xyz)}\Bigg\{ -\log\left(\frac{2xy^3z}{1+x^2}\right) \\
+ \log\Big(
-1-y^2-x(2x+x^3-xy^2+yz+x^2yz-y^3z)+(1+x^2+xyz)p(x,y)
\Big)\Bigg\} \\
+ \frac{zx}{y(z+x(y+xz))}\Bigg\{-\log\left(\frac{2xy^3}{1+x^2}\right) \\
+ \log\Big(
xy(-1-x^2+y^2)-(1+x^2)^2z+(-1+x^2)y^2z + (z+x(y+xz))p(x,y)
\Big)\Bigg\}\, ,
\end{multline}
where $p(x,y) = \sqrt{(1+x^2+y^2)^2-4x^2y^2}$.

In the case of $\Psi^{(1)}(n+k+a+1)$ term we proceed essentially the same way. Indeed, using the integral representations for $\Psi^{(1)}$ and binomial coefficients as above, resuming geometric series in $n$, $a$, $k$  and taking residues in  variables entering integral representations for binomial coefficients we get
\begin{multline}
\mathcal{W}_{6,m}^{\la 1 1 4 4\ra}\Big[\Psi^{(1)}(n+k+a+1)\Big] =
\int_0^1dx_1 \frac{xx_1(x+yz+xz^2)\log x_1}{(x_1-1)(1+x^2x_1)(z+x_1(y+xz)(x+yz))}
\end{multline}
The left integration over $x_1$ is straightforward and gives the following expression
\begin{multline}
\mathcal{W}_{6,m}^{\la 1 1 4 4\ra}\Big[\Psi^{(1)}(n+k+a+1)\Big] = \frac{\pi^2 x (x+yz+xz^2)}{6(1+x^2)(x^2z+(1+y^2)z+xy(1+z^2))} \\
-\frac{x}{y(1+x^2)}\Li_2(-x^2) + \frac{xz}{y(x^2z+(1+y^2)z+xy(1+z^2))}\Li_2 \left(
-\frac{xy+x^2z+y^2z+xyz^2}{z}
\right)
\end{multline}
As a final example of a term in the main contribution $\mathcal{W}_{6,m}^{\la 1 1 4 4\ra}$ let us consider the case of $\left(\Psi^{(0)}(n+k+a+1)+\gamma_E\right)^2$. Again, writing integral representations for polygamma functions and binomial coefficients as above, summing resulting geometric series in $n,a,k$ and taking residues in variables entering integral representations for binomial coefficients we get
\begin{multline}
\mathcal{W}_{6,m}^{\la 1 1 4 4\ra}\Big[\left(\Psi^{(0)}(n+k+a+1)+\gamma_E\right)^2\Big] = \int_0^1dx_1\int_0^1dx_2 \\
\frac{x(x+yz+xz^2)}{(1-x_1)(1-x_2)}\Bigg\{
\frac{1}{(1+x^2)(z+(y+xz)(x+yz))} - \frac{x_1}{(1+x^2x_1)(z+x_1(y+xz)(x+yz))} \\
-\frac{x_2}{(1+x^2x_2)(z+x_2(y+xz)(x+yz))} + \frac{x_1x_2}{(1+x^2x_1x_2)(z+x_1x_2(y+xz)(x+yz))}\Bigg\}
\end{multline}
Now, the integrations in $x_1$ and $x_2$ are straightforward and we finally obtain
\begin{multline}
\mathcal{W}_{6,m}^{\la 1 1 4 4\ra}\Big[\left(\Psi^{(0)}(n+k+a+1)+\gamma_E\right)^2\Big] = \\-\frac{\pi^2 x (x+yz+xz^2)}{6(1+x^2)(x^2z+(1+y^2)z+xy(1+z^2))}
+ \frac{x\log (1+x^2) (-4\log x+3 \log (1+x^2))}{2y(1+x^2)} \\
-\frac{xz}{y(z+(y+xz)(x+yz))}\log\left(
\frac{(y+xz)(x+yz)}{(z+(y+xz)(x+yz))^2}
\right)\log\left(
\frac{z}{z+(y+xz)(x+yz)}
\right) \\
-\frac{xz}{2y(x^2z+(1+y^2)z+xy(1+z^2))}\log\left(
\frac{z}{z+(y+xz)(x+yz)}
\right)\log\left(
z (z+(y+xz)(x+yz))
\right) \\
+\frac{x}{y(1+x^2)}\Li_2\left(
\frac{1}{1+x^2}
\right) - \frac{xz}{y (x^2z+(1+y^2)z+xy(1+z^2))}\Li_2\left(
\frac{z}{xy(1+z^2)+z(1+x^2+y^2)}
\right)
\end{multline}
The evaluation of boundary contribution goes similar to the main one. The different terms in boundary contribution at NLO can be written as
\begin{equation}
\mathcal{W}_{6,b}^{\la 1 1 4 4\ra}\Big[ f_{a,n} \Big] =  \sum_{a=1}^{\infty}\sum_{n=0}^{\infty} (-1)^{a+n}
x^{a+2n+1}y^{a-1}\left(z^a+z^{-a}\right)
\binom{a+n}{n} f_{a,n}
\end{equation}
where $f_{a,n}$ are given by terms in the sum of Eq. \eqref{fboundaryNLO}. Take for example the case with $f_{a,n} = \Psi^{(0)}(n+a+1)+\gamma_E$. Using integral representations for polygamma function and binomial coefficient as before, resuming resulting geometric series in $a, n$ variables and taking integral for binomial coefficient by residues we get
\begin{multline}
\mathcal{W}_{6,b}^{\la 1 1 4 4\ra}\Big[\Psi^{(0)}(n+a+1)+\gamma_E \Big] =\int_0^1dx_1\frac{x^2}{1-x_1}\Bigg\{ \\
\frac{1+2xyz+z^2+x^2(1+z^2)}{(1+x^2)(1+x^2+xyz)(z+x(y+xz))} - \frac{x_1 (1+2xyzx_1+z^2+x^2(1+z^2)x_1)}{(1+x^2x_1)(z+x(y+xz)x_1)(1+x(x+yz)x_1)}\Bigg\}
\end{multline}
The left integration in $x_1$ is straightforward and as a result we obtain
\begin{multline}
\mathcal{W}_{6,b}^{\la 1 1 4 4\ra}\Big[\Psi^{(0)}(n+a+1)+\gamma_E \Big] =
-\frac{2x\log (1+x^2)}{y(1+x^2)} \\
+ \frac{xz}{y(xy+z(1+x^2))}\log\left(
\frac{xy+z(1+x^2)}{z}
\right) + \frac{x\log (1+x^2+xyz)}{y(1+x^2+xyz)}\, .
\end{multline}
The results for all other terms both in main and boundary contributions could be found in accompanying {\it Mathematica} notebook. Gathering all contributions and using symbols\footnote{See Appendix \ref{NMHVsymbols} for more details.} to simplify the resulting expression  we finally get
\begin{multline}
\mathcal{W}_{6}^{\la 1 1 4 4\ra, NLO} = \frac{x}{y}\left(
\frac{1}{1+x^2+xyz}+\frac{z}{z+x(y+xz)}
\right) \\
\times\log\left(
\frac{(1+x^2)z}{x^2(z+(y+xz)(x+yz))}
\right)\log\left(
\frac{(1+x^2)(z+(y+xz)(x+yz))}{y^2z}
\right) \\
+ \frac{x}{y(1+x^2)}\Bigg\{
-\frac{\pi^2}{6} + \log^2x - \log^2\left(
\frac{1+x^2}{xy^2}
\right) + 4\log^2 y + \Li_2\left(
\frac{1}{1+x^2}
\right) + \Li_2\left(
\frac{x^2}{1+x^2}
\right)
\Bigg\} \\
+ \frac{xz}{y (z+(y+xz)(x+yz))}\Bigg\{\frac{\pi^2}{6} -\log^2x
-\log^2\left(
\frac{1+x^2}{x}
\right)-2\log^2 y \\
 + 2\log^2\left(
 \frac{yz}{x(z+(y+xz)(x+yz))}
 \right) - \Li_2\left(
 \frac{1}{1+x^2}
 \right) - \Li_2\left(
 \frac{x^2}{1+x^2}
 \right)\Bigg\}\, ,
\end{multline}
which coincides with $(\mathcal{R}_6^{\la1144\ra}\mathcal{W}_6^{BDS})^{NLO}$, in agreement with \cite{GeneralizedUnitarityN4SYM,NMHV3loop}. See Appendix \ref{Dixonresult} for notation.

\subsection{Prescription for arbitrary order}

The LO and NLO resummation for other NMHV hexagon components\footnote{The starting expressions for our resummation algorithm  similar to Eq.\eqref{W6_1144_start} can be obtained from the results of \cite{CordovaEffectiveParticles}.} goes similar to considered in previous two subsections case of  $\W_6^{\la 1144\ra}$ component. Moreover, it is easy to see, that similar technique is also applicable for one effective particle contribution to MHV hexagon. Indeed, from \cite{MHVresummation} we have
\begin{multline}\label{MHV_pope}
\W_6^{MHV,NLO} =  2\sum_{n=0}^{\infty}\int\frac{du}{2\pi i}
x^{2n+2}y^{2u}\mu_{0,n}^{MHV, non-gluonic}(u) \\
+ \sum_{a=1}^{\infty}\sum_{n=0}^{\infty}\int\frac{du}{2\pi i}
x^{2n+a}y^{2u}(z^a+z^{-a})\left[
\mu_{a,n}^{MHV, gluonic}(u) + x^2\mu_{a,n}^{MHV, non-gluonic}(u)
\right], \\
\end{multline} 
where 
\begin{align}\label{MHV_pope_1}
\mu_{a,n}^{MHV, gluonic}(u) &= \frac{(-1)^{a+n}\Gamma\left(\frac{a}{2}-u\right)}{\left(\frac{a}{2}-u\right)\left(\frac{a}{2}+u\right)\Gamma (n+1)\Gamma (a+n)} \frac{\Gamma\left(n+u+\frac{a}{2}\right)^2}{\Gamma\left(u+\frac{a}{2}\right)}\, , \\
\mu_{a,n}^{MHV, non-gluonic}(u) &= \frac{(-1)^{a+n}\Gamma\left(\frac{a}{2}-u\right)}
{\left(\frac{a}{2}+u\right)\Gamma (n+1)\Gamma (a+n+2)} 
\frac{\Gamma\left(n+u+\frac{a}{2}+1\right)^2}{\Gamma\left(u+\frac{a}{2}+1\right)} \, .
\end{align}
Taking residues in $u=\frac{a}{2}+k$, $k\geq 0$ we get
\begin{multline}\label{MHV_pope_residues}
\W_6^{MHV,NLO} = 2\sum_{n=0}^{\infty}\sum_{k=1}^{\infty} x^{2n+2}y^{2k}\frac{(-1)^{k+n}}{k(n+1)}\binom{n+k}{k}\binom{n+k}{n}   \\
+ \sum_{a,k=1}^{\infty}\sum_{n=0}^{\infty} x^{2n+2+a}y^{2k+a} (z^a+z^{-a})(-1)^{a+k+n}\binom{a+n+k}{k}\binom{a+n+k}{n} \\
\times \left[
\frac{1}{k(n+1)} + \frac{1}{(k+a)(n+a+1)}
\right] + \mbox{boundary terms}
\end{multline}
Staring from this expression for $\W_6^{MHV, NLO}$ we may follows the same steps as for NMHV hexagon in previous subsection. Namely, we   use integral representation for binomial coefficients, express simple fractions $1/k$, $1/(n+1)$, $1/(k+a)$, $1/(n+a+1)$ in terms of polygamma functions and introduce integral representations for the latter. Now summing geometrical series in $a$
, $n$ and $k$ variables we continue with taking residues in variables entering integral representations for binomial coefficients. The left integration in variables entering integral representations of polygamma functions are then  more or less straightforwardly taken in terms of multiple polylogarithms \cite{multiple-polylogs1,multiple-polylogs2}. The same technique should be also applicable for the resummation in the case for polygonal Wilson loops with $n>6$, see \cite{BelitskyResummation1,BelitskyResummation2} for tree level resummation in this case. We also think that the presented technique should be applicable to  the resummation of contributions from several effective particles. However, to be on the save side here we state the algorithm for the resummation of one effective particle contributions to hexagons but for arbitrary loop order of weak coupling expansion.  The necessary steps are given by
\begin{enumerate}
\item Following \cite{CordovaEffectiveParticles} write down the one effective particle contribution to hexagon POPE component you are interested in and expand it in coupling constant up to required loop order. For example, appendix \ref{MeasureEnergiesMomenta} contains corresponding expression in the case of $\W_6^{[2,2]}$ component.
\item Take residues in rapidity of effective particle. It is convenient to first use reflection identities \eqref{reflection-identity-1} and \eqref{reflection-identity-2} to isolate singular terms with known Taylor expansions. It is also useful to transform present polygamma functions to the same argument as far as possible using Eq.\eqref{Psitransform}.
\item Transform the obtained summand to the form of a product of binomial coefficients with simple fractions. For binomial coefficients write down integral representations as in Eq.\eqref{binomial-integral-rep}. In the case of simple fractions express the latter in terms of polygamma functions using Eq.\eqref{simple-fractions-polygamma} and eventually write down integral representations for polygamma functions present.
\item Sum the series present. Now, they are all of geometric progression type and could be easily summed.
\item   Take residues in variables entering integral representations for binomial coefficients.
\item Take integrals in variables entering integral representations of polygamma functions. These are integrals from rational functions and are frequently encountered in calculation of multiloop Feynman diagrams. In particular, they appear in the process of direct integration over Feynman parameters. When, the latter integrals satisfy criterion of linear reducibility \cite{linear-reducibility-1,linear-reducibility-2} one can come with algorithmic way of expressing required integrals  in terms of multiple polylogarithms \cite{multiple-polylogs1,multiple-polylogs2}. In our case there are could be also roots from quadratic polynomials present. The latter however may be rationalized with variable change, see for example \cite{RationalizingRoots}.

\end{enumerate}

\section{Conclusion}

In this paper we presented an algorithmic approach for computing one effective particle contributions to hexagon scattering amplitudes applicable at in principle arbitrary order of perturbation theory. The approach reduces the problem of evaluation of integral over effective particle rapidity and sums over effective particle helicity and descendant number to the problem of evaluation of integrals over rational functions, otherwise known as periods, in terms of multiple polylogarithms \cite{multiple-polylogs1,multiple-polylogs2}. If the latter integrals satisfy the criterion of linear reducibility \cite{linear-reducibility-1,linear-reducibility-2}, then there is an algorithmic way for taking such integrals. In the problem at hand, the integrals may also contain roots of quadratic polynomials. The latter however could be also treated in algorithmic way \cite{RationalizingRoots}.

The presented approach has the potential for the generalization both for higher point scattering amplitudes and contributions with more then one effective particle. This will be the subject of one of our further publications.

\section*{Acknowledgements}

This work was supported by the Foundation for the
Advancement of Theoretical Physics and Mathematics “BASIS”.

\appendix

\section{The remainder for NMHV$_6$ superamplitude}\label{Dixonresult}

The $\mathcal{R}_n^{(k)}$ remainder function is defined to all orders of perturbation theory as the ratio of $\mbox{N}^k\mbox{MHV}_n$ and $\mbox{MHV}_n$ amplitudes:
\begin{equation}
\mathcal{R}_n^{(k)}=\frac{A_n^{(k)}}{A_n^{(0)}}.
\end{equation}
In NMHV case $k=1$. From now on we will drop the $(k)$ superscript and stick with NMHV$_6$ case only. Note, that due to universal (independent from particles helicities) structure of IR divergences the remainder function is IR finite. In addition, it is also dual conformal invariant.

Using momentum twistors  $\mathcal{Z}_i = (\vll_i ,\mu_i , \vle_i)$ \cite{MomentumTwistors} and splitting the $\mathcal{R}_6$ remainder function into $even$ and $odd$ parts we have \cite{GeneralizedUnitarityN4SYM,NMHV3loop}:
\begin{equation}
\mathcal{R}_6=\mathcal{R}_6^{even}+\mathcal{R}_6^{odd} ,
\end{equation}
where
\begin{eqnarray}
\mathcal{R}_6^{even}&=&
\frac{[13456]+[12346]}{2}V(u_1,u_2,u_3)+
\frac{[12456]+[12345]}{2}V(u_2,u_3,u_1)\nonumber\\
&+&\frac{[23456]+[12356]}{2}V(u_3,u_1,u_2),
\end{eqnarray}
and\footnote{It is convenient to define different set of arguments for $\tilde{V}$, which, however, can be expressed through $u_1,u_2,u_3$ \cite{NMHV3loop}. Since we are actually will be interested only in $V$ function we will not write them here.}
\begin{eqnarray}
\mathcal{R}_6^{odd}&=&([12346]-[13456])\tilde{V}(u_1,u_2,u_3)+([12456]-[12345])\tilde{V}(u_2,u_3,u_1)\nonumber\\
&+&([23456]-[12356])\tilde{V}(u_3,u_1,u_2).
\end{eqnarray}
$V$ and $\tilde{V}$ are scalar functions, which  depend only on (dual)conformal cross ratios and coupling constant $g$. $[abcde]$ is dual conformal invariant (five-bracket) defined as
\begin{eqnarray}
[i \; j \; k \; l \; m] = \frac{\delta^4 (\la i \; j \; k \; l\ra\vle_m + \text{cyclic permutation})}{\la i \; j \; k \; l\ra\la j \; k \; l \; m\ra\la k \; l \; m \; i\ra\la l \; m \; i \; j\ra\la m \; i \; j \; k\ra}
\end{eqnarray}
with four-brackets $\la i \; j \; k \; l\ra$ being defined through bosonic components of momentum twistors $Z_i =  (\vll_i ,\mu_i )$  as
\begin{eqnarray}
\la i \; j \; k \; l\ra = \varepsilon_{A B C D} Z_i^A Z_j^B Z_k^C Z_l^D = det(Z_iZ_jZ_kZ_p),
\end{eqnarray}
The expansion of functions $V$ and $\tilde{V}$ in coupling constant reads
\begin{eqnarray}
V(u_1,u_2,u_3)&=&1+\sum_{l=1}(2g^2)^lV^{(l)}(u_1,u_2,u_3)\\
\tilde{V}(u_1,u_2,u_3)&=&\sum_{l=1}(2g^2)^l\tilde{V}^{(l)}(u_1,u_2,u_3).
\end{eqnarray}
All information about helicity content of remainder function is contained in $[abcde]$ rational functions, which are
all loop exact. All coupling constant dependence is through $V$ and $\tilde{V}$ functions only. Note also,
that due to the six term identity
\begin{eqnarray}
[23456]-[13456]+[12456]-[12356]+[12346]-[12345]=0
\end{eqnarray}
at leading order we have
\begin{eqnarray}
\mathcal{R}_6^{LO} = [12345]+[12356]+[13456],
\end{eqnarray}
which is $[1,2\ra$ BCFW representation of normalized tree level six point amplitude.

Dual conformal cross ratios for six point functions can be conveniently written in terms of dual
variables\footnote{$x_{ij}^2=\left(\sum_{k=i}^{j-1}p_k\right)^2$ with $p_i$ standing for momentum of $i$'th particle and
	sum being understood in cyclic sense.} as
\begin{eqnarray}
u_1\equiv v=\frac{x_{13}^2x_{46}^2}{x_{14}^2x_{36}^2},~u_2\equiv w=\frac{x_{24}^2x_{51}^2}{x_{25}^2x_{41}^2},~u_3\equiv u=\frac{x_{35}^2x_{62}^2}{x_{36}^2x_{52}^2}.
\end{eqnarray}
Using the relation $x_{jk}^2 = \frac{\la j-1, j, k-1, k\ra}{\la j-1, j\ra\la k-1, k\ra}$ the latter could be also written in terms of four-brackets
\begin{equation}
u = \frac{\la 1236\ra\la 3456\ra}{\la 2356\ra\la 1346\ra}, \quad
v =\frac{\la 1234\ra\la 1456\ra}{\la 1245\ra\la 1346\ra},\quad  w =\frac{\la 1256\ra\la 2345\ra}{\la 1245\ra\la 2356\ra}.
\end{equation}
At next-to-leading order  $V^{(l)}$ function is given by:
\begin{equation}
V^{(1)}(u_1,u_2,u_3)=\frac{1}{2}\left(\sum_{i=1}^3\Li_2(u_i)+\Big(\log (u_1)+\log (u_3)\Big)\log (u_2)-\log (u_1)\log (u_3)
-\frac{\pi^2}{3}\right),
\end{equation}
while $\tilde{V}^{(1)}=0$, i.e. there is no contribution to $\mathcal{R}_6^{(1)odd}$ at NLO.

As an illustration of our summation method we have chosen a particular component of $\mathcal{R}_6$
function proportional to $\eta_1\eta_1\eta_4\eta_4$ Grassmann monomial: $\mathcal{R}_6^{\la 1144\ra}$. At NLO  it is given by
\begin{eqnarray}
2\mathcal{R}_6^{\la 1144\ra,NLO}=2\mathcal{R}_6^{even,NLO}\Big|_{\eta_1\eta_1\eta_4\eta_4}&=&g^{2}([13456]+[12346])\Big|_{\eta_1\eta_1\eta_4\eta_4}
V^{(1)}(v,w,u)+\nonumber\\
&+&
g^{2}([12456]+[12345])\Big|_{\eta_1\eta_1\eta_4\eta_4}
V^{(1)}(w,u,v). \nonumber \\
\end{eqnarray}
The coefficients in front of $V^{(1)}(v,w,u)$ and $V^{(1)}(w,u,v)$ are then given by:
\begin{eqnarray}
\Big([13456]+[12346]\Big)\Big|_{\eta_1\eta_1\eta_4\eta_4}&=&
\frac{\la 1356 \ra\la 3456 \ra}{\la 1345 \ra\la 1456 \ra\la 1346 \ra}+
\frac{\la 1236 \ra\la 2346 \ra}{\la 1234 \ra\la 1346 \ra\la 1246 \ra},\\
\Big([12456]+[12345]\Big)\Big|_{\eta_1\eta_1\eta_4\eta_4}&=&
\frac{\la 1256 \ra\la 2456 \ra}{\la 1245 \ra\la 1456 \ra\la 1246 \ra}+
\frac{\la 1235 \ra\la 2345 \ra}{\la 1234 \ra\la 1345 \ra\la 1245 \ra}.
\end{eqnarray}
At LO this leads to
\begin{equation}
\mathcal{R}_6^{LO}\Big|_{\eta_1\eta_1\eta_4\eta_4} =
\frac{\la 2345 \ra\la 1235 \ra}{\la 1234 \ra\la 1345 \ra\la 1245 \ra}+
\frac{\la 3456 \ra\la 1356 \ra}{\la 1345 \ra\la 1456 \ra\la 1346 \ra}.
\end{equation}

In collinear OPE approach the kinematics for six point amplitude is parameterized by three real parameters: ${\tau,~\sigma,~\phi}$. Dual conformal cross ratios $u,v,w$ as well as
all $\la abcd \ra$ invariants are then expressed via these parameters using explicit parametrization
of hexagon momentum twistors (here we use notation from the main text $x=e^{-\tau}, y=e^{\sigma}, z=e^{i\phi}$):
\begin{eqnarray}
\begin{pmatrix}
Z_1   \\
Z_2   \\
Z_3   \\
Z_4   \\
Z_5   \\
Z_6
\end{pmatrix}=
\begin{pmatrix}
yz^{-1/2} & 0 & z^{1/2}x^{-1}& xz^{1/2} \\
1 & 0 & 0& 0  \\
-1 & 0 & 0& 1  \\
0 & 1 & -1& 1  \\
0 & 1 & 0& 0  \\
0 & y^{-1}z^{-1/2} & x^{-1}z^{1/2}& 0
\end{pmatrix}
\end{eqnarray}
For example for dual conformal cross ratios we get:
\begin{eqnarray}
u&=&\frac{z}{x y + (1 + x^2 + y^2) z + x y z^2}\\
v&=&\frac{y^2 z}{(1 + x^2) (x y + (1 + x^2 + y^2) z + x y z^2)}\\
w&=&\frac{x^2}{1 + x^2}
\end{eqnarray}
while the coefficients in front of $V^{(1)}(v,w,u)$ and $V^{(1)}(w,u,v)$ functions take the form
\begin{eqnarray}
\Big([13456]+[12346]\Big)\Big|_{\eta_1\eta_1\eta_4\eta_4}&=&
\frac{x^3 z (x + y z)}{(-x z - x^3 z - x^2 y z^2) (x^2 y + x z + x^3 z + x y^2 z + x^2 y z^2)}\nonumber\\
&+&
\frac{-x z (x y^2 z + x^2 y z^2)}{y (x y + z + x^2 z) (x^2 y + x z + x^3 z + x y^2 z + x^2 y z^2)}
\nonumber\\
\Big([12456]+[12345]\Big)\Big|_{\eta_1\eta_1\eta_4\eta_4}&=&
\frac{x^2 z}{(-z - x^2 z) (x y + z + x^2 z)}+
\frac{-x^3 z^3}{(-z - x^2 z) (-x z - x^3 z - x^2 y z^2)}.\nonumber\\
\end{eqnarray}
In this parametrization  the limit $x \rightarrow 0$ (large $\tau$) describes regime when momenta $p_1$ and $p_6$ are becoming collinear.

The LO contribution to remainder function in terms of collinear OPE variables reads:
\begin{eqnarray}
\mathcal{R}_6^{LO}\Big|_{\eta_1\eta_1\eta_4\eta_4}&=&\frac{x}{y}\left(
\frac{z}{z+(y+xz)(x+yz)} - \frac{1}{1+x^2}\,
\right).
\end{eqnarray}

Within collinear OPE approach one actually computes not the reminder function $\mathcal{R}_6$ itself,
but another finite function $\mathcal{W}_6$ of the same dual conformal invariants, which is related to $\mathcal{R}_6$ as
\begin{equation}
\mathcal{R}_6=\frac{\mathcal{W}_6}{\mathcal{W}_6^{MHV}},
\end{equation}
where  $\mathcal{W}_6^{MHV}=\mathcal{R}_6^{MHV}\mathcal{W}_6^{BDS}$. Here  $\mathcal{R}_6^{MHV}$ is $\mbox{MHV}_6$ remainder function and $\mathcal{W}_6^{BDS}$ is known function of cusp anomalous dimension
\begin{eqnarray}
\Gamma_{cusp}(g)=4g^2-\frac{4\pi^2}{3}g^4+O(g^6)
\end{eqnarray}
and dual conformal invariants $u_i$:
\begin{eqnarray}
\mathcal{W}_6^{BDS}(u_1,u_2,u_3)&=&\exp\Big\{\frac{\Gamma_{cusp}(g)}{4}\Big(\Li_2(u_2)-\Li_2(1-u_1)-\Li_2(1-u_3)\nonumber\\
&+&\log^2(1-u_2)-\log (u_1)\log (u_3)+\log (u_1/u_3)\log (1-u_2)+\frac{\pi^2}{6}\Big)\Big\}.\nonumber\\
\end{eqnarray}
At NLO $\mathcal{R}_6^{MHV}=1$ and we are left with the following relation between collinear OPE result and the NMHV amplitude remainder function:
\begin{equation}
\mathcal{R}_6^{\la1144\ra,NLO}=\left(\frac{\mathcal{W}_6^{\la1144\ra}}{\mathcal{W}_6^{BDS}(u,w,v)}\right)^{NLO},
\end{equation}
where it is assumed that $\mathcal{W}_6^{\la1144\ra}/\mathcal{W}_6^{BDS}$ should be expanded up to $O(g^2)$.

\section{Measures, energies and momenta}\label{MeasureEnergiesMomenta}

The expression for charged pentagon component $\W_6^{[2,2]}$ considered in the main body of the paper written in terms of a sum over effective particles contributions is given by \cite{CordovaEffectiveParticles}:
\begin{multline}
\W_6^{[2,2]} = \sum_{\Phi}\int\frac{du}{2\pi} e^{E_{\Phi}(u)\tau + i p_{\Phi}(u)\sigma + i m_{\Phi}\phi}\mu_{\Phi}^{[2,2]}(u) \\
= \sum_{n=0}^{\infty}\sum_{a=-\infty}^{\infty}\int\frac{du}{2\pi}
e^{-E_{a,n}^{eff}(u)\tau + i p_{a,n}^{eff}(u)\sigma + i a\phi}\mu_{a,n}^{[2,2], eff}(u)\, ,
\end{multline}
where energies and momenta of effective particles have the form
\begin{align}
E_{a,n}^{eff}(u) = 2n+1+|a|+4g \left(\Q\M\cdot\kappa_{a,n}^{eff}\right)_1\, ,\quad
p_{a,n}^{eff}(u) = 2u - 4g\left(\Q\M\cdot\tilde{\kappa}_{a,n}^{eff}\right)_1
\end{align}
Here, infinite matrices $\Q$ and $\M$ are given by \cite{Basso:2014koa_III}:
\begin{equation}
\Q_{ij} = \delta_{ij}(-1)^{i+1}i ,\quad \M = \left[\mathbb{I}+\mathbb{K}\right]^{-1} ,\quad \mathbb{K}_{ij} = 2j(-1)^{j(i+1)}\int_0^{\infty}\frac{dt}{t}\frac{J_i(2gt)J_j(2gt)}{e^t-1} .
\end{equation}
Up to NLO we have
\begin{equation}
\Q\M =
\begin{pmatrix}
1-\frac{g^2\pi^2}{3} & -4g^3\zeta(3) \cr -4g^3\zeta(3) & -2+\frac{2g^4\pi^4}{15}
\end{pmatrix}+O(g^4),
\end{equation}
The infinite vectors $\kappa_{a,n}^{eff}$ and $\tilde{\kappa}_{a,n}^{eff}$ are build from Bethe string describing effective particle transforming in vector representation of $SU(4)$ and labeled by helicity $a$ and descendant number $n$. This way we get  \cite{CordovaEffectiveParticles}:
\begin{align}
\kappa_{a,n}^{eff} &=  k_a(u) + \sum_{j=1}^{n+2}\kappa_{\psi_S}(
u - i(
\frac{|a|-3}{2}+j)) + \sum_{j=1}^n \kappa_{\psi_S} (u - i (\frac{|a|+1}{2}+j)) , \\
\tilde{\kappa}_{a,n}^{eff} &=  \tilde{k}_a(u) + \sum_{j=1}^{n+2}\tilde{\kappa}_{\psi_S}(
u - i(
\frac{|a|-3}{2}+j)) + \sum_{j=1}^n \tilde{\kappa}_{\psi_S} (u - i (\frac{|a|+1}{2}+j)) ,
\end{align}
where \cite{Basso:2014koa_III}:
\begin{align}
\kappa_a (u) &\equiv (\kappa_{a,1}(u), \kappa_{a,2}(u), \ldots) , \quad \kappa_{\psi_S}(u)\equiv (\kappa_{\psi_S, 1}, \kappa_{\psi_S, 2}, \ldots) , \nonumber \\
\tilde{\kappa}_a (u) &\equiv (\tilde{\kappa}_{a,1}(u), \tilde{\kappa}_{a,2}(u), \ldots) , \quad \tilde{\kappa}_{\psi_S}(u)\equiv (\tilde{\kappa}_{\psi_S, 1}, \tilde{\kappa}_{\psi_S, 2}, \ldots)
\end{align}
with ($J_j(z)$ are Bessel functions)
\begin{align}
\kappa_{a,j}(u) &= \int_0^{\infty}\frac{dt}{t (e^t-1)} J_j (2gt)\left(
J_0(2gt) - \cos (ut) e^{f_t (j,a)}\right) , \\
\tilde{\kappa}_{a,j}(u) &= (-1)^{j+1}\int_0^{\infty}\frac{dt}{t(e^t-1)} J_j (2gt)\sin (ut) e^{f_t (j+1,a)} , \\
\kappa_{\psi_S,j}(u) &= \frac{(-1)^{j/2}(1+(-1)^j)}{4j}\left(
\frac{g}{x(u)}
\right)^j , \quad \tilde{\kappa}_{\psi_S, j} = \frac{(-1)^{\frac{j+1}{2}}(1-(-1)^j)}{4j}\left(
\frac{g}{x(u)}
\right)^j .
\end{align}
Here $x(u)$ is Zhukovsky variable $x(u) = \frac{1}{2}(u+\sqrt{u^2-4g^2})$ and $f_t (j,a) = t(1-\frac{|a|-(-1)^j}{2})$

The measures for effective particles are also build on the basis of their Bethe string representations and are given by \cite{CordovaEffectiveParticles}:
\begin{equation}
\mu_{a,n}^{[2,2],eff} = g^{-1} \frac{M_{a,n}(u)}{f_{a,0}(u) f_{a,0}(-u)}\exp_{a,n}^{eff} (u) ,
\end{equation}
where
\begin{align}
\exp_{a,n}^{eff}(u) &= \exp\left[
-2(\kappa_{a,n}^{eff})^t\cdot \Q\M\cdot \kappa_{a,n}^{eff}
+ 2 (\tilde{\kappa}_{a,n}^{eff})^t\cdot\Q\M\cdot\tilde{\kappa}_{a,n}^{eff}
\right] \\
\log\left(
f_{a,0}(u)
\right) &= \int_0^{\infty}\frac{dt}{t (e^t-1)}\left(
J_0(2gt)-1
\right)\left[
\frac{1}{2}J_0(2gt) + \frac{1}{2} - e^{\frac{(1-|a|)t}{2}-iut}
\right]
\end{align}
and ($x^{[a]} = x(u-i a/2)$)
\begin{equation}
M_{a,n}(u) = \frac{M_{a,0}(u)}{\Gamma (n+1)\Gamma (|a|+n+1)}\prod_{l=1}^n \left(
x^{[2l+|a|+1]}
\right)^2
\end{equation}
\begin{multline}
M_{a,0}(u) =g (-1)^a\Gamma\left(
iu+\frac{|a|+1}{2}\right)\Gamma\left(
-iu+\frac{|a|+1}{2}
\right) \\
\times\frac{x^{[1+|a|]}}{x^{[1-|a|]}}\frac{x^{[1-|a|]}x^{[1+|a|]}-g^2}{\sqrt{\left(x^{[1-|a|]}\right)^2-g^2}\sqrt{\left(
x^{[1+|a|]}		
\right)^2-g^2}}
\end{multline}

\section{Simplifying $\mathcal{W}_6^{\la 1 1 4 4\ra}$ with symbols}\label{NMHVsymbols}

To compare the result of pentagon OPE resummation\footnote{It can be found in accompanying {\it Mathematica} notebook.} for $\mathcal{W}_6^{\la 1 1 4 4\ra}$ with the known results from generalized unitarity and bootstrap \cite{GeneralizedUnitarityN4SYM,NMHV3loop} we need to simplify our expression. The most convenient way to do it it to use symbols technique \cite{symbols1,symbols2,symbols3}, in particular the {\it Mathematica} package \texttt{PolyLogTools} \cite{PolyLogTools}. In fact, we only need the following two symbols:
\begin{align}
\Li_2 (z) &\to -(1-z)\otimes z , \\
\log (x)\log(y) &\to x\otimes y + y\otimes x .
\end{align}
Note, that symbol mapping is blind to constants\footnote{The constants could be fixed by comparing expressions at some fixed kinematical point.} and satisfy the relations
\begin{align}
a_1\otimes\ldots\otimes a_ia_j\otimes\ldots \otimes a_n &= a_1\otimes\ldots\otimes a_i\otimes\ldots \otimes a_n + a_1\otimes\ldots\otimes a_j\otimes\ldots \otimes a_n \\
a_1\otimes\ldots\otimes a_i^n\otimes\ldots\otimes a_n &= n (a_1\otimes\ldots\otimes a_i\otimes\ldots\otimes a_n) .
\end{align}

To simplify consideration we will consider the simplification of the difference of the our resulting expression with \cite{GeneralizedUnitarityN4SYM,NMHV3loop}. In the case of 1 loop NMHV$_6$ amplitude contrary to the case of 1 loop MHV amplitude \cite{MHVresummation} the resulting expressions contains rational factors in front dilogarithms and logarithms. The latter after partial fraction in $x$ variable are given by
\begin{align}
p_1 &= \frac{x}{(1+x^2)y} , \quad p_2 = \frac{xz}{y(xy+z+zx^2)} , \quad
p_3 = \frac{x}{y(1+x^2+xyz)} , \nonumber \\
p_4 &= \frac{xz}{y(xy+z+x^2z)(z^2-1)} , \quad p_5 = \frac{xz^3}{y(xy+z+x^2z)(z^2-1)} , \\
p_6 &= \frac{x}{y(1+x^2+xyz)(z^2-1)}, \quad p_7 = \frac{xz}{y(xy+z+x^2z+y^2z+xyz^2)} .\nonumber
\end{align}
The usage of symbol map with \texttt{PolyLogTools} package reduces to the application of just three commands \texttt{SymbolMap}, \texttt{SymbolExpand} and \texttt{SymbolFactor} together with the simplification of symbol entries with {\it Mathematica} command \texttt{FullSimplify}. Using symbol map for the considered difference it easy to show that coefficients in front of $p_1$ and $p_7$ rational factors  are equal to zero, while the coefficient in front of $p_2$ equal to the coefficients in front of $p_3$, $p_4$ factors and minus coefficient in front of $p_5$. Taking into account found functional identities and using again partial fractioning in $x$ variable it is easy to see that the coefficient in front of $p_6$ in the expression for $\mathcal{W}_6^{\la 1 1 4 4\ra}$ also cancels.  This finishes the proof of equivalence of our and \cite{GeneralizedUnitarityN4SYM,NMHV3loop} results.

\bibliographystyle{hieeetr}
\bibliography{refsOPE}

\end{document}